\documentclass[superscriptaddress,aps,twocolumn,longbibliography,prl]{revtex4-1}
\usepackage{amsmath,amssymb}
\usepackage{graphicx}
\usepackage{braket}
\usepackage{color}
\usepackage{epsfig}
\usepackage{epstopdf}
\usepackage{hyperref}
\usepackage{bibentry}
\usepackage{array,booktabs,multirow}
\hypersetup{
colorlinks = true,
linkcolor = [rgb]{0.7,0.,0.},
citecolor = [rgb]{0.25, 0.41, 0.88},
urlcolor = [rgb]{0.25, 0.41, 0.88}}

\newcolumntype{V}{>{\centering\arraybackslash} m{.4\linewidth} }

\graphicspath{{figures/}}

\begin{document}
\title{ Anisotropy as a diagnostic test for distinct tensor network wavefunctions of integer and half-integer spin Kitaev quantum spin liquids }

\author{Hyun-Yong Lee}
\affiliation{Department of Applied Physics, Graduate School, Korea University, Sejong 30019, Korea}
\affiliation{Division of Display and Semiconductor Physics, Korea University, Sejong 30019, Korea}

\author{Takafumi Suzuki}
\affiliation{Graduate School for Engineering, University of Hyogo, Himeji, Hyogo 670-2280, Japan}

\author{Yong Baek Kim}
\email{ybkim@physics.utoronto.ca}
\affiliation{Department of Physics, University of Toronto, Toronto, Ontario M5S 1A7, Canada}

\author{Naoki Kawashima}
\email{kawashima@issp.u-tokyo.ac.jp}
\affiliation{Institute for Solid State Physics, University of Tokyo, Kashiwa, Chiba 277-8581, Japan}

\date{\today}

\begin{abstract}
	Contrasting ground states of quantum magnets with the integer and half-integer spin moments are the manifestation of many-body quantum interference effects. In this work, we investigate the distinct nature of the integer and half-integer spin quantum spin liquids in the framework of the Kitaev’s model on the honeycomb lattice. The models with arbitrary spin quantum numbers are not exactly solvable in contrast to the well-known quantum spin liquid solution of the spin-1/2 system. We use the tensor network wavefunctions for the integer and half-integer spin quantum spin liquid states to unveil the important difference between these states. We find that the distinct sign structures of the tensor network wavefunction for the integer and half-integer spin quantum spin liquids are responsible for completely different ground states in the spatially anisotropic limit. Hence the spatial anisotropy would be a useful diagnostic test for distinguishing these quantum spin liquid states, both in the numerical computations and experiments on real materials. We support this discovery via extensive numerics including the tensor network, DMRG, and exact diagonalization computations.
\end{abstract}
\maketitle

{\it Introduction - } Recently there have been immense experimental and theoretical efforts to unveil a quantum spin liquid state in frustrated magnets with bond-dependent interactions, which include $\alpha$-RuCl$_3$\cite{Khaliullin2005, Jackeli2009, Plumb2014, Sears15, Johnson15, Kim15, Kim16, Yadav16, Zhou17, Banerjee2016, Luke16, Sinn16, Winter16, Leahy17a, Trebst2017, Banerjee17, Catuneanu18, Gohlke18, Winter18, Banerjee2018, Balz19, Wang19} and various polymorphs of Li$_2$IrO$_3$\cite{Chaloupka15, Rau16, Williams16, Perreault15, Katukuri16, Breznay17, Rousochatzakis18, Majumder18}. These activities are largely motivated by the prospect of realizing the Kitaev’s spin-1/2 model on the honeycomb lattice, which allows an exact solution of the quantum spin liquid\cite{Kitaev2006}. Moreover, a number of candidate materials for the spin-1 and spin-3/2 analogs have also been proposed\cite{ Peter19, Inhee20, Xu20}. Given that the integer and half-integer spin models often support different kinds of quantum ground states, it is interesting to explore whether there is any fundamental difference between the integer and half-integer spin quantum spin liquid phases.

In this work, we investigate distinct signatures of the integer and half-integer spin quantum spin liquid states via the tensor network wavefunctions and other numerical tools. In contrast to the spin-1/2 model, the higher-spin Kitaev models are not exactly solvable. On the other hand, there exist numerical studies of the $S\!=\!1$ Kitaev model supporting the existence of a quantum spin liquid ground state\cite{Oitmaa18,Koga18,Peter19,HY19b, Dong19,Khait20,Zhu20}. In particular, an earlier study proposes the tensor network wavefunction for the $S\!=\!1$ Kitaev quantum spin liquid\cite{HY19b}. Here we present the tensor network wavefunctions for arbitrary integer spin quantum number and contrast its properties with those of the spin-1/2 wavefunction. It is shown that the spatial anisotropy in the exchange interactions in the Kitaev model can be used to uncover important differences between the tensor network wavefunctions of the integer and half-integer spin moments. It has been known that the spatially anisotropic limit of the Kitave model for the $S\!=\!1$ and $S\!=\!1/2$ systems leads to the trivial product state and Toric code topological state respectively\cite{Tetsuya20, Kitaev2006}. 

We find that this phenomenon can be understood as a result of different sign structures of the tensor network wavefunctions of the integer and half-integer spin systems. We explicitly demonstrate that the tensor network wavefunction of the integer spin systems allow the phase transition to the trivial product state in the anisotropic limit while the non-trivial sign structure of the half-integer spin tensor network wavefunction is the obstruction to form a trivial product state. This contrasting behavior is generic and represents an important difference between the integer and half-integer spin Kitaev quantum spin liquids. 


%
\begin{figure}[h!]
	\includegraphics[width=0.43\textwidth]{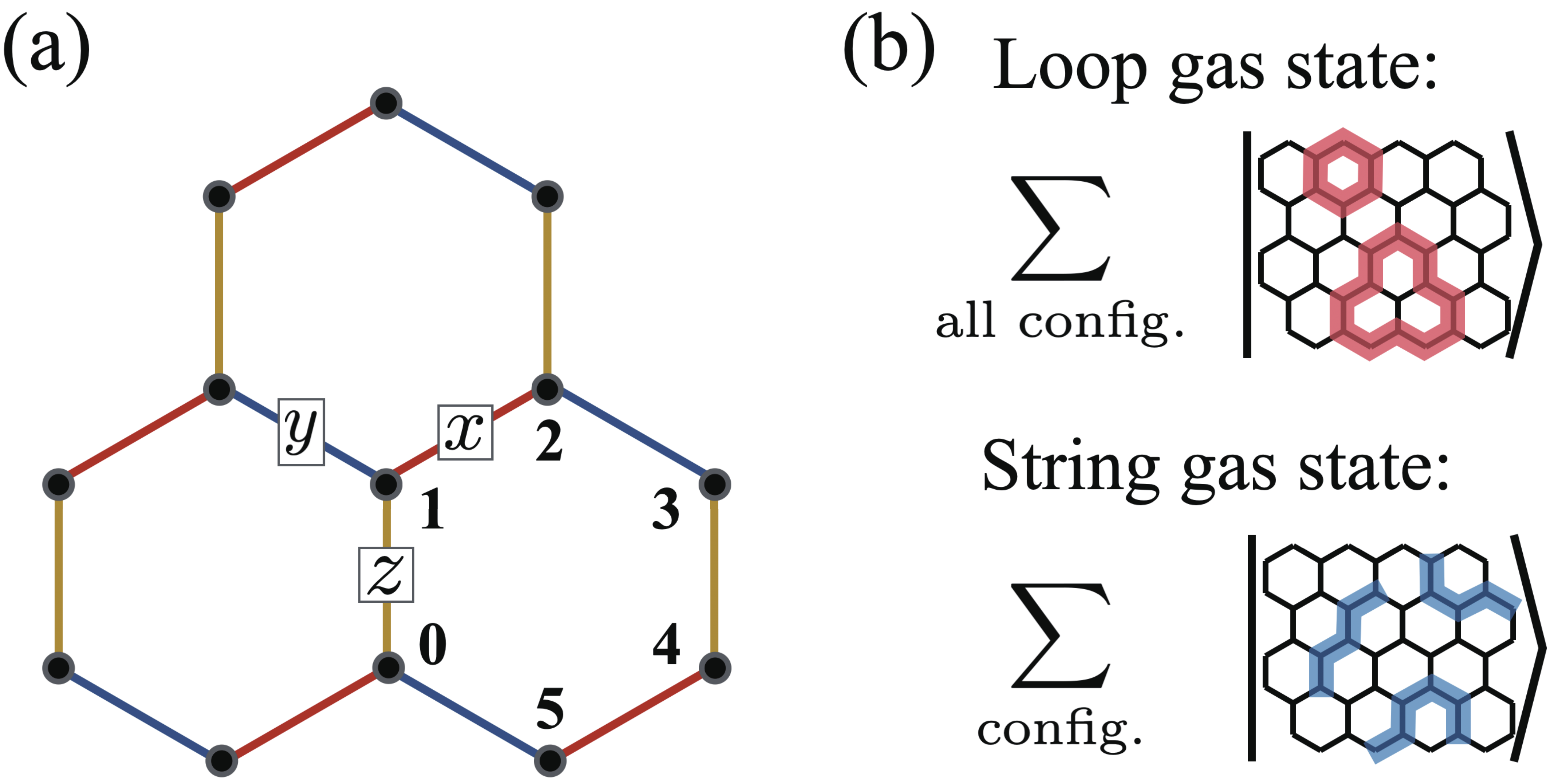} 
	\caption{ (a) Graphical representation of the Kitaev honeycomb model where the red, blue and yellow bonds denote the $x$, $y$ and $z$ bonds, respectively. (b) Schematic figure of the loop gas and string gas states that we utilize to find the ground states.  } 
	\label{fig:schematic}
\end{figure}

{\it Model - } The Hamiltonian of the spin-$S$ Kitaev honeycomb model reads $H = \sum_{\langle ij \rangle_\gamma} H_{ij}^\gamma$ with  
%
	$H_{ij}^\gamma = - K_\gamma S_i^\gamma S_j^\gamma$, 
%
where $\langle ij \rangle_\gamma$ denotes the nearest-neighbor pair $i$ and $j$ sites  on the $\gamma$-bond, and $S^\gamma$ is the spin-$S$ operator with $\gamma = x,y,z$. The model possesses a set of local conserved quantities, the so-called flux $w_p=\pm 1$, which is defined on every hexagon plaquette\,($p$). It can be detected by the flux operator $W_p = U_0^x U_1^y U_2^z U_3^x U_4^y U_5^z $ where $U_j^\gamma = e^{i\pi S_j^\gamma}$, and sites 0-5 are shown in Fig.\,\ref{fig:schematic}. Note that the ground state of the spin-$1/2$ Kitaev model belongs to the vortex-full sector in the current definition of the flux operator, i.e., $\{w_p = -1\}$\cite{Kitaev2006}. On the other hand, that of the spin-$1$ model is in the vortex-free sector $\{w_p=+1\}$\cite{Baskaran08, Koga18, HY19b}. One can define a projector operator projecting any quantum state into a desired flux sector\,($\mathcal{S}$): $\prod_p (1 + (-)^{ l_{\mathcal{S},p} } W_p )$ where $l_{\mathcal{S},p} = 0,1$ determines the flux number at plaquette $p$ depending on the desired sector $\mathcal{S}$, e.g., $l_{\mathcal{S},p} = 0\,(1)$ for all $p$ if the target sector is the vortex-free\,(vortex-full). Expanding the projector operator, it can be recast as the summation over all possible loop configurations of product of $U_i^x,U_i^y$ and $U_i^z$ along the loops. Since all configurations are equally weighted, we refer the projector as `loop gas' \,(LG) operator: $Q_{\rm LG}^{\mathcal{S}}$. Taking into account this structure, the LG operators for the spin-1/2 and spin-1 models were recast as the bond dimension $D=2$ tensor network\,(TN) in Refs.\,\cite{HY19,HY19b}, respectively. Applying $Q_{\rm LG}^{\mathcal{S}}$ to a product state generates the so-called LG state as illustrated in Fig.\,\ref{fig:schematic}\,(b), which serves a great trial wavefunction to simulate the ground state of Kitaev models\cite{HY19,HY20a,HY19b}. Note that the physical and topological properties of the LG state depends on the initial product state that the $Q_{\rm LG}^{\mathcal{S}}$ is applied to\cite{HY20a}.

{\it Loop Gas Operator - } Here, we generalize the LG operator to general integer and half-integer spins. The local tensor $Q_{ijk}$ of the TN operator is defined as 
\begin{align}
	&Q_{000} = \mathbb{I}_{2S+1},
	&&Q_{011} = \zeta\,U^x,\nonumber\\ 
	&Q_{101} = \zeta\,U^y,
	&&Q_{110} = \zeta\,U^z,
	\label{eq:q_tensor}
\end{align}
where $\zeta$ is an extra phase factor which is the unity for integer spin while $-i$ for half-integer spin. Note that the non-trivial phase cannot be eliminated by a gauge transformation, and it plays a key role determining the non-trivial entanglement structure of the half-integer spin model in the strong anisotropic limit as shown below. Contracting the TN with $Q_{ijk}$, one obtains the vortex-full projector $Q^{\rm full}_{\rm LG} = \prod_p (1-W_p)$ for half-integer spin and vortex-free projector $Q^{\rm free}_{\rm LG} = \prod_p (1+W_p)$ for integer spin. See Supplemental Material at [URL will be inserted by publisher] for more details. Using the $Z_2$ gauge symmetry, one can easily obtain $Q_{\rm LG}^{\rm full}$ for integer spin and $Q_{\rm LG}^{\rm free}$ for half-integer spin by inserting a proper tensor in the TN as shown in Table.\,\ref{tab:projector}. In a similar way, one can construct $Q_{\rm LG}^{\mathcal{S}}$ for an arbitrary flux sector $\mathcal{S}$ by decorating the TN of $Q_{\rm LG}^{\rm full/free}$ for each case.

\begin{table}[h!]
  \begin{center}
    \begin{tabular}{|c|c|c|}
      \hline
      spin & integer & half-integer\\
      \hline
       \multirow{-5}{*}{$Q_{\rm LG}^{\rm free}$} & \includegraphics[width=0.2\textwidth]{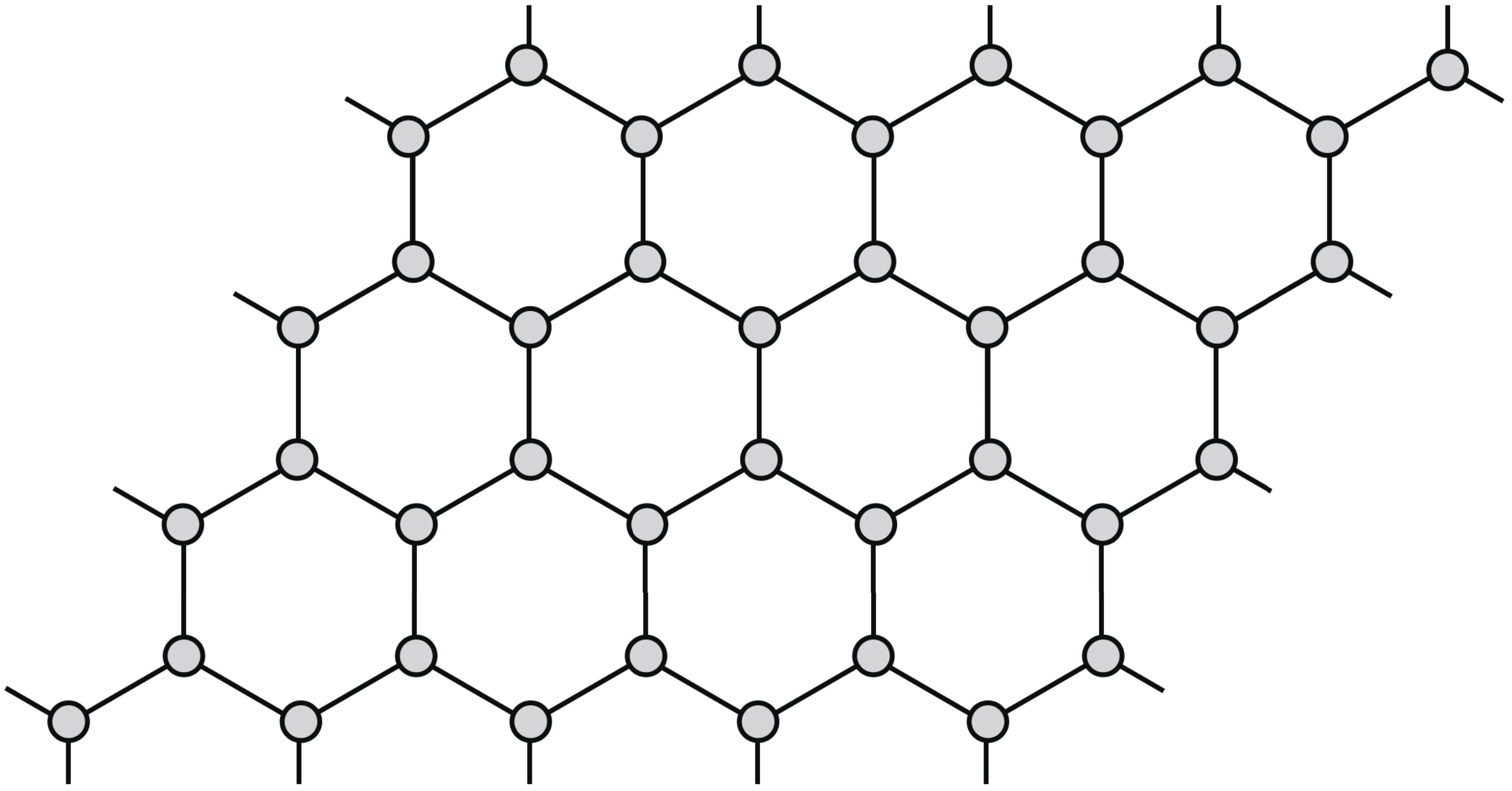} & \includegraphics[width=0.2\textwidth]{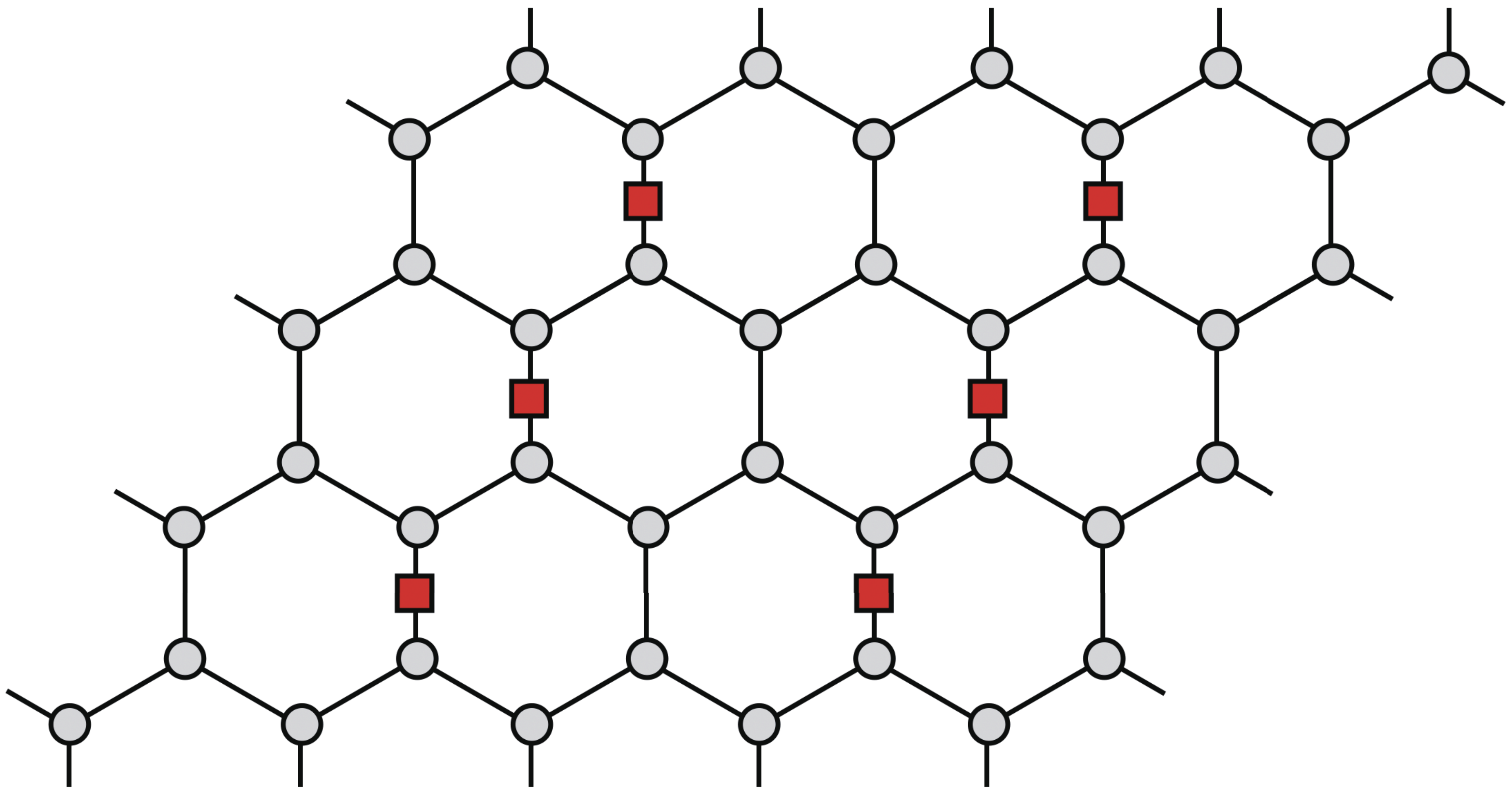}\\
      \hline
      \multirow{-5}{*}{$Q_{\rm LG}^{\rm full}$} &  \includegraphics[width=0.2\textwidth]{decorated.eps} & \includegraphics[width=0.2\textwidth]{bare.eps}\\
         \hline
    \end{tabular}
  \end{center} 
  \caption{Graphical TN representation of the projector $Q_{\rm LG}^{\rm free/full}$ for integer and half-integer spins. Here, the gray circle stands for the $Q$-tensor defined in Eq.\,\eqref{eq:q_tensor}, and the red square for the Pauli matrix $\sigma^z$. }
  \label{tab:projector}
\end{table}
%


{\it Strong Anisotropic Limit - } In the strong anisotropic limit\,(say $K_z=1$ and $K_x,K_y \rightarrow 0$), the Hamiltonian becomes $H_z = -\sum_{\langle ij \rangle_z} S_i^z S_j^z$. Then, using $[H_z, Q_{\rm LG}^{\mathcal{S}}]=0$, one can easily verify that a wavefunction $|\psi_z^{\mathcal{S}}\rangle = Q_{\rm LG}^{\mathcal{S}} |\!\uparrow\rangle $ becomes the exact (degenerate) ground state of $H_z$ regardless of the spin magnitude $S$, where $|\!\uparrow\rangle \! = \! \otimes_i |\!\uparrow_i\rangle $ stands for the product state of fully polarized magnetic state aligned in the $z$-direction, i.e., $S^z \!=\! +S$. See Supplemental Material at [URL will be inserted by publisher] for more details. Therefore, one can always find the exact ground state at the strong anisotropic point in the $D=2$ TN representation regardless of spin-$S$. Now, we show how the extra phase $\zeta$ in the $Q$-tensor affects the resulting state. To this end, we first note that the local state $|\!\uparrow_j\rangle$ is transformed under the action of $U^\gamma_j$ as follows:
%
	$U_j^x |\!\uparrow_j\rangle = e^{i\pi S} |\!\downarrow_j\rangle$,
	$U_j^y |\!\uparrow_j\rangle = (-)^{2S} |\!\downarrow_j\rangle$ and
	$U_j^z |\!\uparrow_j\rangle = e^{i\pi S} |\!\uparrow_j\rangle$.
%
Then, let us apply a loop operator generated by the LG operator to $|\uparrow\rangle$ as illustrated below:

\begin{align}
	\includegraphics[width=0.43\textwidth]{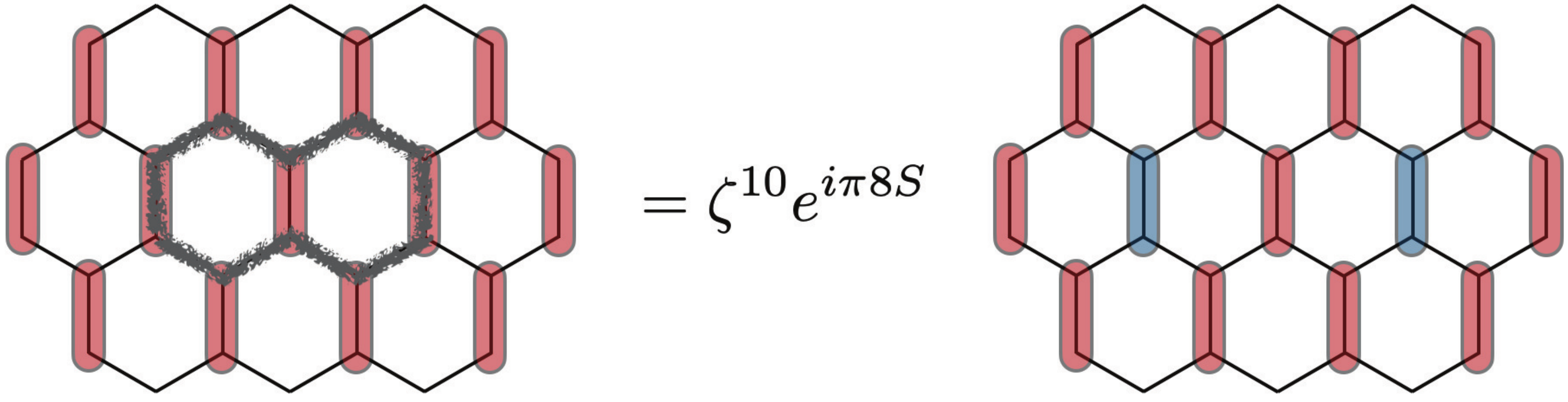}.
\end{align}
Here, the filled ellipses stand for effective spin-1/2's, i.e., red: $|\tilde{\uparrow}_i\rangle = |\uparrow_i \uparrow_{i+\hat{z}}\rangle$, blue: $|\tilde{\downarrow}_i\rangle = |\downarrow_i \downarrow_{i+\hat{z}}\rangle$, and the thick gray line denotes the loop operator. Note that the loop operators flip some of the effective spins and generate an overall phase factor $\zeta^L e^{i \pi 4n S}$ depending on its length $L$ and shape of the loop determining an integer $n$. The phase factor $e^{i \pi 4n S}$ is the unity regardless of the (original) spin magnitude. It denotes that, in the case of integer spin\,($\zeta=1$), $|\psi_z^{\rm free}\rangle = Q_{\rm LG}^{\rm free}|\tilde{\uparrow}\rangle$ consists of all kinds of effective spin-up/down configurations with the same phase in the thermodynamic limit. Since there is no preferred direction of the loop\,[Eq.\,\eqref{eq:q_tensor}], the probability that a $z$-bond is occupied by a loop is half. Consequently, the resulting state in the thermodynamic limit is recast as $ |\psi_z^{\rm free}\rangle = \bigotimes_{i=1}^{N_z} \frac{1}{\sqrt{2}} ( |\tilde{\uparrow}_i\rangle + |\tilde{\downarrow}_i\rangle )$ with $N_z$ being the number of $z$-bonds. This can be directly checked by computing the overlap between $Q_{\rm LG}^{\rm free}|\uparrow\rangle$ and $\bigotimes_{i=1}^{N_z} \frac{1}{\sqrt{2}} ( |\tilde{\uparrow}_i\rangle + |\tilde{\downarrow}_i\rangle )$ with the proper normalization. To see this, we first note that the norm of the LG state is equivalent to the number of configurations of the eight vertex model\,($Z_{\rm 8ver}$) on an effective square lattice obtained by combining two sublattices on the $z$-bond: 
\begin{align}
	\langle \psi^{\mathcal{S}} | \psi^{\mathcal{S}} \rangle = \langle \uparrow | (Q_{\rm LG})^2 |\uparrow \rangle = Z_{\rm 8ver} \langle \uparrow | Q_{\rm LG} |\uparrow \rangle = Z_{\rm 8ver}.	\nonumber
\end{align}
Here, we use the facts that $Q_{\rm LG}^{\mathcal{S}}$ is hermitian for the first equality, the product of two loop configurations leads to another loop configuration for the second equality, and that a loop operator\,(products of $W_p$) flips some of up-spins to down-spins for the last equality. See Supplemental Material at [URL will be inserted by publisher] for more details. Then, taking into account the normalization, one can directly evaluate the overlap:
\begin{align}
	\frac{1}{(Z_{\rm 8ver} 2^{N_z})^{\frac{1}{2}}} \bigotimes_{i=1}^{N_z} ( \langle \tilde{\uparrow}_i | + \langle \tilde{\downarrow}_i | ) Q_{\rm LG}^{\rm free} |\uparrow\rangle = \left(\frac{Z_{\rm 8ver}}{2^{N_z}}\right)^{\frac{1}{2}},
\end{align}
where we use the fact that all $Z_{\rm 8ver}$ configurations expanded by $Q_{\rm LG}^{\rm free}$ are realized by  $\bigotimes_{i=1}^{N_z} ( | \tilde{\uparrow}_i \rangle + | \tilde{\downarrow}_i \rangle )$. Since the entropy per site of the eight vertex model in the thermodynamic limit is $\log 2$\cite{Baxter07}, the normalization factor becomes $Z_{\rm 8ver}=2^{N_z}$, and thus the overlap is the  unity. In a similar way, one can verify that the (degenerate) ground states in other flux sectors can be recast as trivial product states as well. Thus, the $Z_2$ gauge symmetry of $Q_{\rm LG}^{\mathcal{S}}$ is redundant when it is applied to $|\tilde{\uparrow}\rangle$ in the case of integer spin. In brief, the ground state of integer spin model in the strong anisotropic limit is a simple product state that is consistent with Ref.\,\cite{Tetsuya20}. 

On the other hand, in the case of half-integer spin, the non-trivial phase $\zeta^L$ prevents $|\psi_z^{\mathcal{S}}\rangle = Q_{\rm LG}^{\mathcal{S}} |\tilde{\uparrow}\rangle$ from being a simple product state irrespective of the flux sector. To be more concrete, the plaquette operator $W_p$ acting on $|\uparrow\rangle$ is identical to a plaquette operator $\widetilde{W}_p = \tilde{\sigma}^y_l \tilde{\sigma}^z_u \tilde{\sigma}^y_r \tilde{\sigma}^z_d $ acting on $|\tilde{\uparrow}\rangle$ as depicted below:
\begin{align}
	\includegraphics[width=0.4\textwidth]{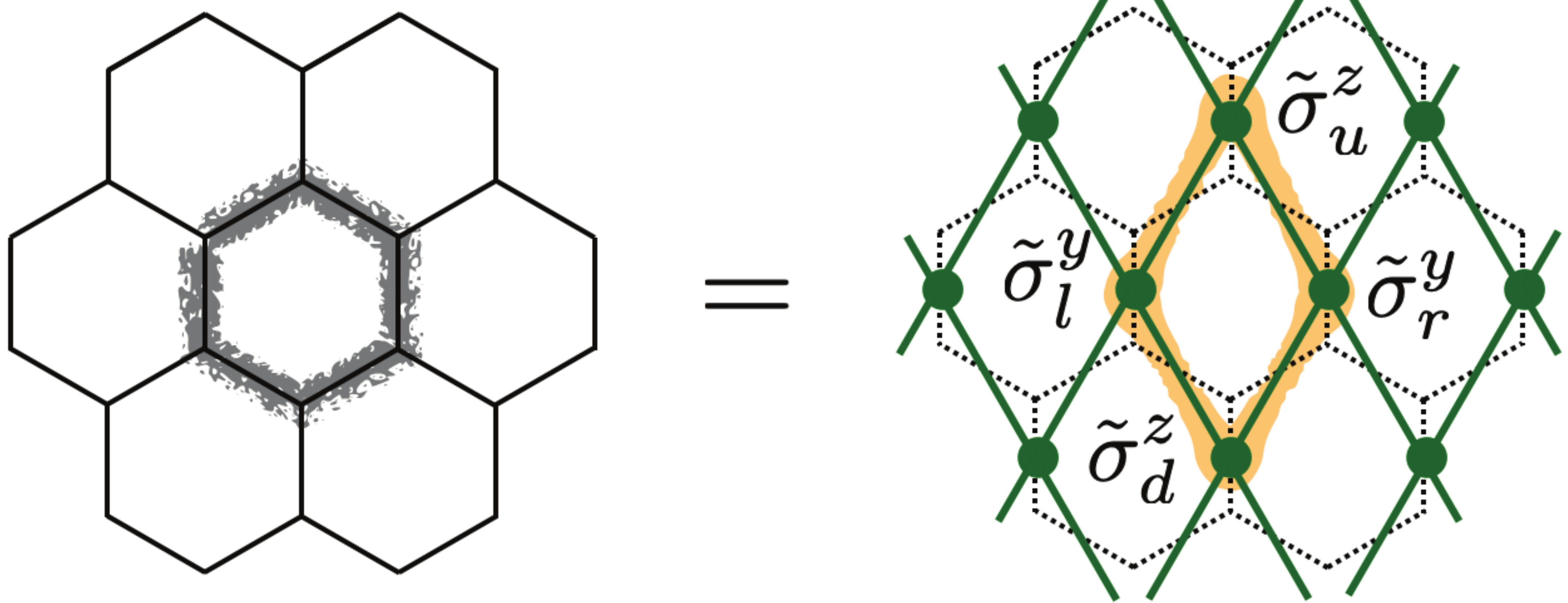}.\nonumber
\end{align}
Therefore, $|\psi_z^{\rm free}\rangle = Q_{\rm LG}^{\rm free}|\uparrow\rangle$ is identical to $\prod_p ( 1 + \widetilde{W}_p ) |\tilde{\uparrow}\rangle$ on the effective square lattice that is the ground state of the Hamiltonian $\widetilde{H} = - \sum_p \widetilde{W}_p$, i.e., the effective Hamiltonian near the strong anisotropic limit derived by Kitaev in Ref.\,\cite{Kitaev2006}. The effective model can be unitarily transformed into the Toric code\cite{Kitaev2006}, and thus $|\psi_z^{\rm free}\rangle$ hosts the $Z_2$ topological order. This applies to higher (half-integer) spins identically, and we therefore conclude that the ground state of the half-integer spin Kitaev models in the strong anisotropic limit is the $Z_2$ spin liquid. It indicates that the ground state phase diagrams in terms of spatial anisotropy of the integer and half-integer spin Kitaev model are qualitatively different from each other. Since the gauge symmetry cannot be spontaneously broken\cite{Elitzur75, Batista05}, the LG picture obtained in the strong anisotropic limit will survive over the whole phase diagram irrespective of spin-$S$. In the case of half-integer spin, as shown above, the LG hosts the long-range entanglement even at the strong anisotropic limit, and thus the topologically non-trivial ground states are guaranteed against the anisotropy. On the other hand, in the case of integer spin, the non-trivial phase or KSL phase may be fragile against the anisotropy in that the LG state in the strong anisotropic limit and can be smoothly connected to the product state. In order to confirm the validity of this argument, we present numerical results on the spin-1 model and discuss its phase diagram below.

\begin{figure}[t!]
	\includegraphics[width=0.49\textwidth]{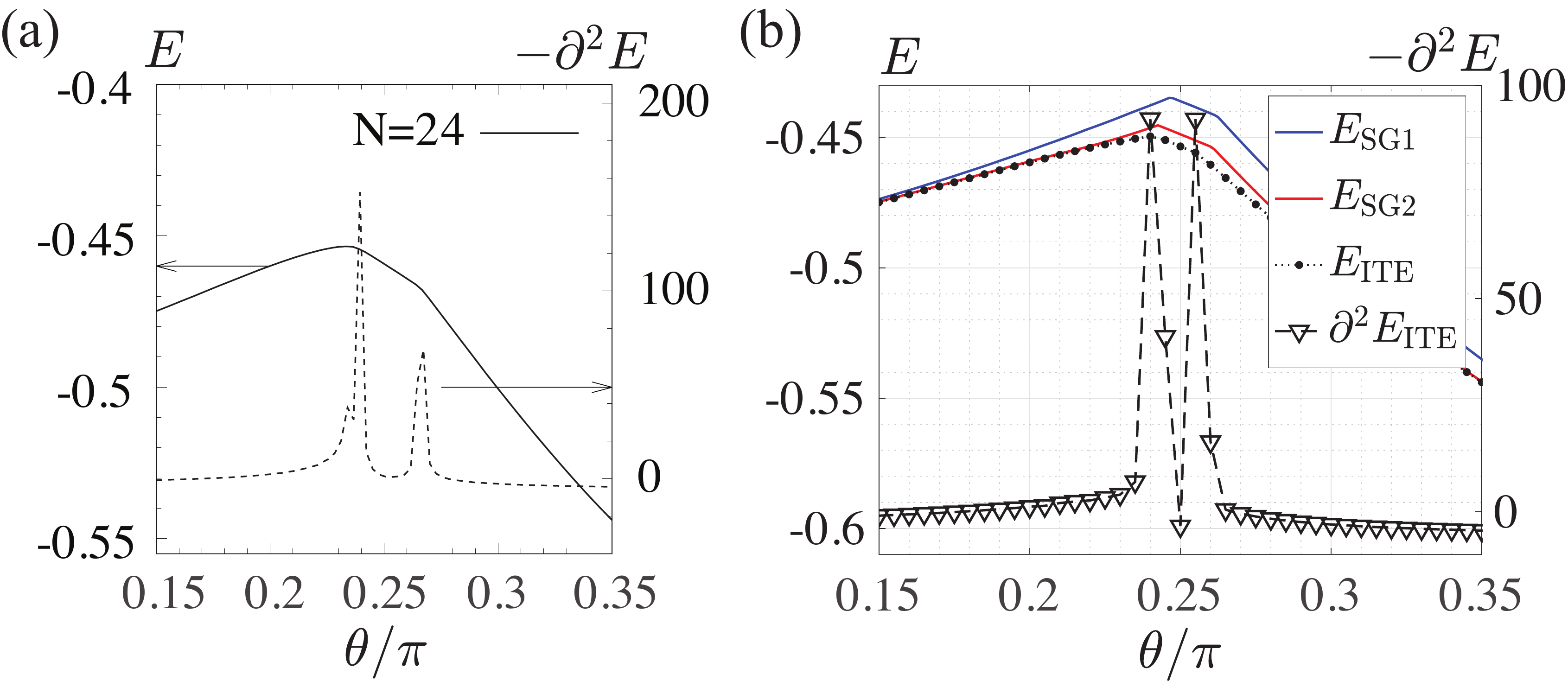} 
	\caption{ (a) Ground state energy and its second derivative of 24-site cluster obtained by ED (b) Variational energies obtained by the first and second order SG ansatz and ITE as a function of the model parameter $\theta$. The sharp peaks in the second derivative of the energy\,(black solid line) imply that two phase transitions occur around the isotropic point $\theta$=$\pi/4$.} 
	\label{fig:energy_compare}
\end{figure}

\begin{figure}[t!]\includegraphics[width=0.25\textwidth]{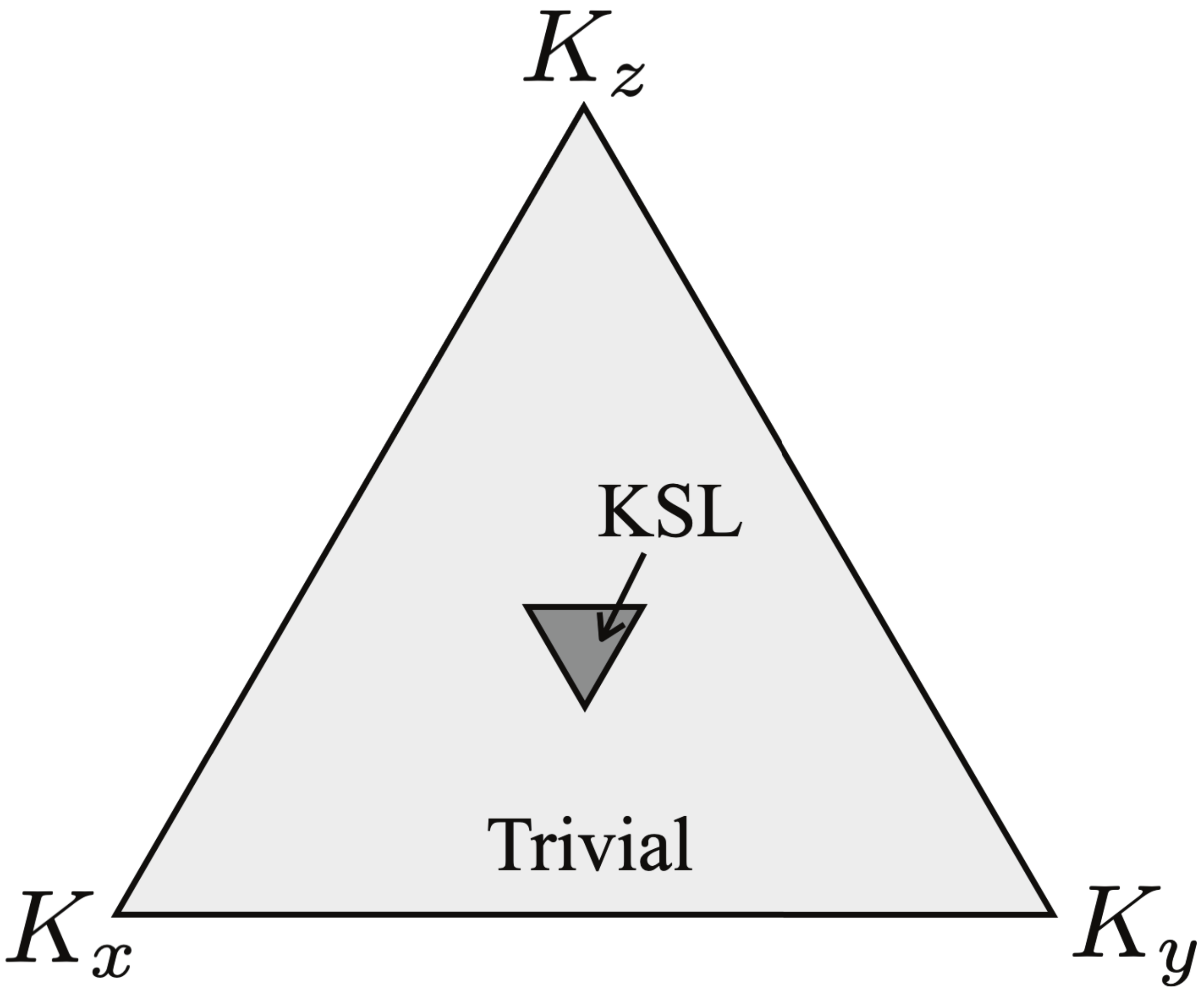} 
	\caption{ Schematic phase diagram of the spin-1 Kitaev model, where the trivial phase is smoothly transformed to the product state at each strong anisotropic point. } 
	\label{fig:phase_diagram}
\end{figure}

{\it Phase Diagram of Spin-1 model - } To carve out the phase diagram as a function of the anisotropy, we employ the ED, DMRG and TN approaches. The Lanczos method\cite{Lanczos} is utilized for ED, and the TN of the infinite system is optimzed with two different schemes, i.e., the imaginary time evolution\,(ITE)\cite{Tao08} and variational wavefunction approach\cite{HY19}. In Ref.\,\cite{HY19}, it was shown that generating open-ended loop, or string configurations as depicted in Fig.\,\ref{fig:schematic}\,(b) is useful to lower the variational energy, while the physical properties including gauge structure and symmetry are intact. We refer the `dressed' LG wavefunction as the string gas\,(SG) wavefunction. 
The anisotropy of the model is parameterized as follows:
%
	$K_x \!=\! K_y \!=\! \sin\theta$ and $K_z \!=\! \cos \theta$.
%
%
\begin{figure}[t!]
	\includegraphics[width=0.48\textwidth]{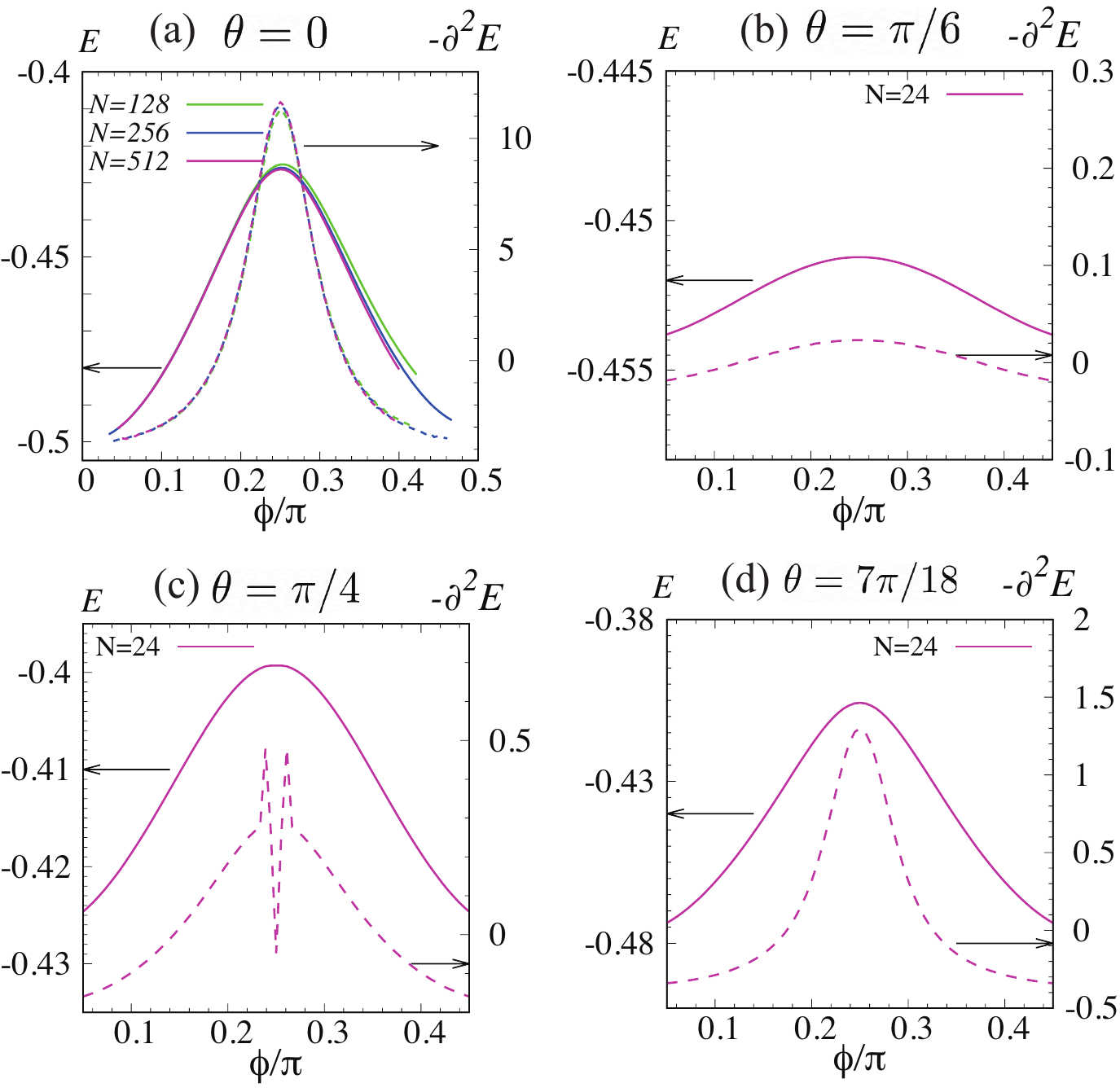} 
	\caption{ Ground state energy and its second derivative. The anisotropy of the model is parametrized as $(K_x,K_y,K_z) = (\sin\theta\cos\phi, \sin\theta\sin\phi, \cos\phi)$ (a) The chain limit and (b)-(d) intermediate regions.} 
	\label{fig:interacting_chain}
\end{figure}
It was conjectured that, in the semiclassical limit, the ground states of the Kitaev honeycomb model live in the vortex-free sector with higher spin-$S$\cite{Baskaran08}. In order to check its validity for the $S\!=\!1$ quantum model, we have performed the ED calculations on a variety of clusters and system sizes and then confirmed that the ground states are in the vortex-free sector over the entire range $0\leq\theta\leq\pi/2$. 
The system size dependence of the results and analysis on the flux sector are presented in details in Supplemental Material at [URL will be inserted by publisher]. Based on that, we optimize the TN wavefunction in the vortex-free sector utilizing $Q_{\rm LG}^{\rm free}$. The energy density\,($E$) and its second derivative\,($\partial^2_\theta E$) obtained by ED and TN are presented in Fig.\,\ref{fig:energy_compare}. It is worth noting that the two-parameter SG wavefunction\,(SG1) provides reasonable variational energy compared to the ITE optimization only near the strong anisotropic limit, while the three-parameter SG wavefunction\,(SG2) gives competitive energy throughout the phase diagram\,[Fig.\,\ref{fig:energy_compare}\,(b)]. Therefore, the ground state of the $S=1$ Kitaev model can be efficiently described by the SG wavefunction for arbitrary $(K_x,K_y,K_z)$. Both ED and TN ansatz find two first-order phase transitions near the isotropic point\,($\theta=\pi/4$), at which the ground state is the Kitaev spin liquid\,(KSL)\cite{HY19b,Dong19,Zhu20,Khait20}. Note that the KSL is stable only in a narrow window. In other words, it is not as robust as the one of the spin-1/2 model with respect to the anisotropy. We find that the KSL phase is surrounded by a trivial phase smoothly connected to the trivial product states at each strong anisotropic point. In the chain limit\,($\theta\!=\!\pi/2$ or $K_z\!=\!0$), we introduce another anisotropy parameter, say $\phi$, such that $K_x \!=\! \cos\phi$ and $K_y \!=\! \sin\phi$. Interestingly, the DMRG simulation finds no signature of transition in $0\leq \phi \leq \pi/2$ as shown in Fig.\,\ref{fig:interacting_chain}\,(a). The energy and its second derivative are featureless without system size dependence. It indicates that a ground state at the strong anisotropic point can be smoothly deformed into another without passing throughout a transition. This is one of the characteristics  of the spin-1 model distinguished from the spin-half model where the strong anisotropic limits are separated by a quantum phase transition at $\phi=\pi/4$\cite{Kitaev2006}. Turning on $\theta$ slightly, i.e., weakly interacting chains, the phase diagram is featureless yet as shown in Fig.\,\ref{fig:interacting_chain}\,(b). Increasing further the inter-chain interaction\,($K_z$), the KSL appears near $\theta=\pi/4$\,[Fig.\,\ref{fig:interacting_chain}\,(c)] and disappears again as it approaches the dimer limit\,[Fig.\,\ref{fig:interacting_chain}\,(d)]. 
Based on the exact result in the strong anisotropic limit and extensive numerical results, we suggest a schematic phase diagram of the spin-1 Kitaev model in Fig.\,\ref{fig:phase_diagram}. The gapped or gapless nature of the KSL near the isotropic point is not completely clear yet. However, it is certain that the KSL phase is fragile and thus survives only in a small region. We also speculate that the phase diagram is valid even for higher integer spin Kitaev models.

{\it Conclusions -} In this letter, we have provided the TN wavefunctions of the spin-$S$ Kitaev quantum spin liquids and investigated the difference between the integer and half-integer spin systems in the anisotropic limit. First, we have shown that the so-called LG operator $Q^{\mathcal{S}}_{\rm LG}$, which can be efficiently written in terms of the $D \!=\! 2$ TN, maps a particular reference state to the exact ground state in the strong anisotropic limit. Further, it has been rigorously shown that the topological nature of the ground states depends only on the quantum number of the spin, i.e., integer or half-integer. The integer spin LG state becomes a simple product state while that of the half-integer spin sustains the long-range entanglement, leading to the $Z_2$ topological order, i.e. the Topic code, regardless of the flux sector and magnitude of the spin. Therefore, in the case of half-integer spin, the non-trivial topological feature remains throughout the phase diagram as a function of the anisotropy. On the other hand, in the case of integer spin, the trivial phase may take a large portion of the phase diagram, and the Kitaev spin liquid is stable only in a small region near the isotropic point. Using extensive numerical calculations including ED, DMRG and 2D tensor networks, we have carved out the phase diagram of the anisotropic spin-1 Kitaev model. We have confirmed that the ground states are in the vortex-free sector throughout the phase diagram as a function of the anisotropy and further found that the strong anisotropic limits are adiabatically connected each other. It has been also shown that indeed the trivial state occupies a large portion of the phase diagram and the $S\!=\!1$ Kiatev spin liquid state is confined to an area near the isotropic point. The sensitivity of the integer spin Kitaev spin liquids and the robustness of the half-integer cases would be an important diagnostic tool for future numerical studies of more general theoretical models and experiments on real materials.

{\it Acknowledgements -} A part of computation in the present work is executed on computers at the Supercomputer Center, ISSP, University of Tokyo and at the Research Center for Nano-Micro Structure Center, University of Hyogo. H.-Y.L. was supported by a Korea University Grant and National Research Foundation of Korea\,(NRF-2020R1I1A3074769). T.S.’s work is supported by Kinoshita Research Foundation. Y.B.K. is supported by the NSERC of Canada and the Killam Research Fellowship of the Canada Council for the Arts. N.K.'s work is financially supported by MEXT Grant-in-Aid for Scientific Research (B) (19H01809). This research was supported by the MEXT project ``Exploratory Challenge on Post- K computer''(Frontiers of Basic Science: Challenging the Limits).

H.-Y. Lee and T. S. contributed equally to this work.

\bibliographystyle{apsrev}
\bibliography{reference.bib}

\clearpage
\onecolumngrid
\begin{center}
\textbf{\large  Supplemental Material}
\end{center}

\setcounter{equation}{0}
\setcounter{figure}{0}
\setcounter{table}{0}
\setcounter{page}{1}

\author{Hyun-Yong Lee}
\affiliation{Institute for Solid State Physics, University of Tokyo, Kashiwa, Chiba 277-8581, Japan}
\affiliation{Department of Applied Physics, Graduate School, Korea University, Sejong 30019, Korea}
\affiliation{Division of Display and Semiconductor Physics, Korea University, Sejong 30019, Korea}

\author{Takafumi Suzuki}
\affiliation{Graduate School for Engineering, University of Hyogo, Himeji, Hyogo 670-2280, Japan}

\author{Yong Baek Kim}
\email{ybkim@physics.utoronto.ca}
\affiliation{Department of Physics, University of Toronto, Toronto, Ontario M5S 1A7, Canada}

\author{Naoki Kawashima}
\email{kawashima@issp.u-tokyo.ac.jp}
\affiliation{Institute for Solid State Physics, University of Tokyo, Kashiwa, Chiba 277-8581, Japan}

\section{Spin-$S$ Loop gas operator in the tensor network representation}

In this section, we discuss the LG operator, $Q_{\rm LG}$, in the tensor network\,(TN) representation. The Hamiltonian of the Kitaev model of general spin-$S$ reads

\begin{align}
	H = -K_x \sum_{\langle ij \rangle_x} S_i^x S_j^x 
	-K_y \sum_{\langle ij \rangle_y} S_i^y S_j^y	
	-K_z \sum_{\langle ij \rangle_z} S_i^z S_j^z,
\end{align}
where $\langle ij \rangle_\gamma$ denotes the nearest neighboring sites $i$ and $j$ on the $\gamma$-bond, and $S^\gamma$ is the spin-$S$ operator. The flux operator commuting with the Hamiltonian can be defined as $W_p = U_1^x U_2^y U_3^z U_4^x U_5^y U_6^z$ with the local spin-rotation operator $U_j^\gamma = e^{i\pi S_j^\gamma}$. Then, the flux operator satisfies the following commutation relations regardless of $S$: $\{U^\gamma, S^{\gamma'} \} = 0$ for $\gamma\neq\gamma'$, $[U^\gamma,S^\gamma]=0$, and thus $[W_p, H]=0$ irrespective of the choice of $K_x$, $K_y$ and $K_z$. In addition, the flux operators on different plaquettes commute each other, i.e., $[W_p, W_{p'}] = 0$ regardless of spin-$S$. On the other hand, the local spin rotation operator $U^\gamma$ satisfies the following commutation relation: 

\begin{itemize}
	\item Integer spin: $U^\gamma U^{\gamma'} = \epsilon_{\gamma \gamma' \gamma''}^2 \times U^{\gamma''}$
	\item Half-integer spin: $U^\gamma U^{\gamma'} = - \epsilon_{\gamma \gamma' \gamma ''} \times U^{\gamma''}$ for $\gamma \neq \gamma'$, \quad $(U^\gamma)^2 = -1$,
\end{itemize}
where $\epsilon_{ijk}$ is the Levi-Civita symbol.
This difference leads to the fundamental distinction between the integer spin and half-integer spin LG states in the anisotropic limit. 

\subsection{Integer spin}

The TN representation of the $S=1$ loop gas\,(LG) operator defined in Ref.\,\cite{HY19b} can be generalized to that of integer spins, and its local tensor, $Q_{ijk}$, is given by

\begin{align}
	Q_{000}	= \mathbb{I}_{2S+1},\quad
	Q_{011}	= U^x,\quad
	Q_{101}	= U^y,\quad
	Q_{110}	= U^z,
\end{align}
where the dimension of the virtual indices is two, i.e., $i,j,k = 0,1$, and $\mathbb{I}_{2S+1}$ denotes the $(2S+1)$-dimensional identity operator. Using $[U^\gamma,U^{\gamma'}]=0$, one can verify the following relations 

\begin{align}
	U^x Q_{ijk} = \sigma^x_{jj'} \sigma^x_{kk'} Q_{ij'k'},\quad
	U^y Q_{ijk} = \sigma^x_{kk'} \sigma^x_{ii'} Q_{i'jk'},\quad
	U^z Q_{ijk} = \sigma^x_{ii'} \sigma^x_{jj'} Q_{i'j'k}.
	\label{eq:Q_integer}
\end{align}
The above relation was discussed in Ref.\,\cite{HY19b} for the case of spin-one but holds for all integer spins. As shown in Ref.\,\cite{HY19b}, the above relation allows us to verify that the TN operator made of $Q_{ijk}$, say $Q_{\rm LG}$, is identical to $\prod_p(1+W_p)$, i.e., the $Q_{\rm LG}$ operator projects any quantum state into the vortex-free sector.

\subsection{Half-integer spin}

\begin{figure}[h!]
	\includegraphics[width=0.46\textwidth]{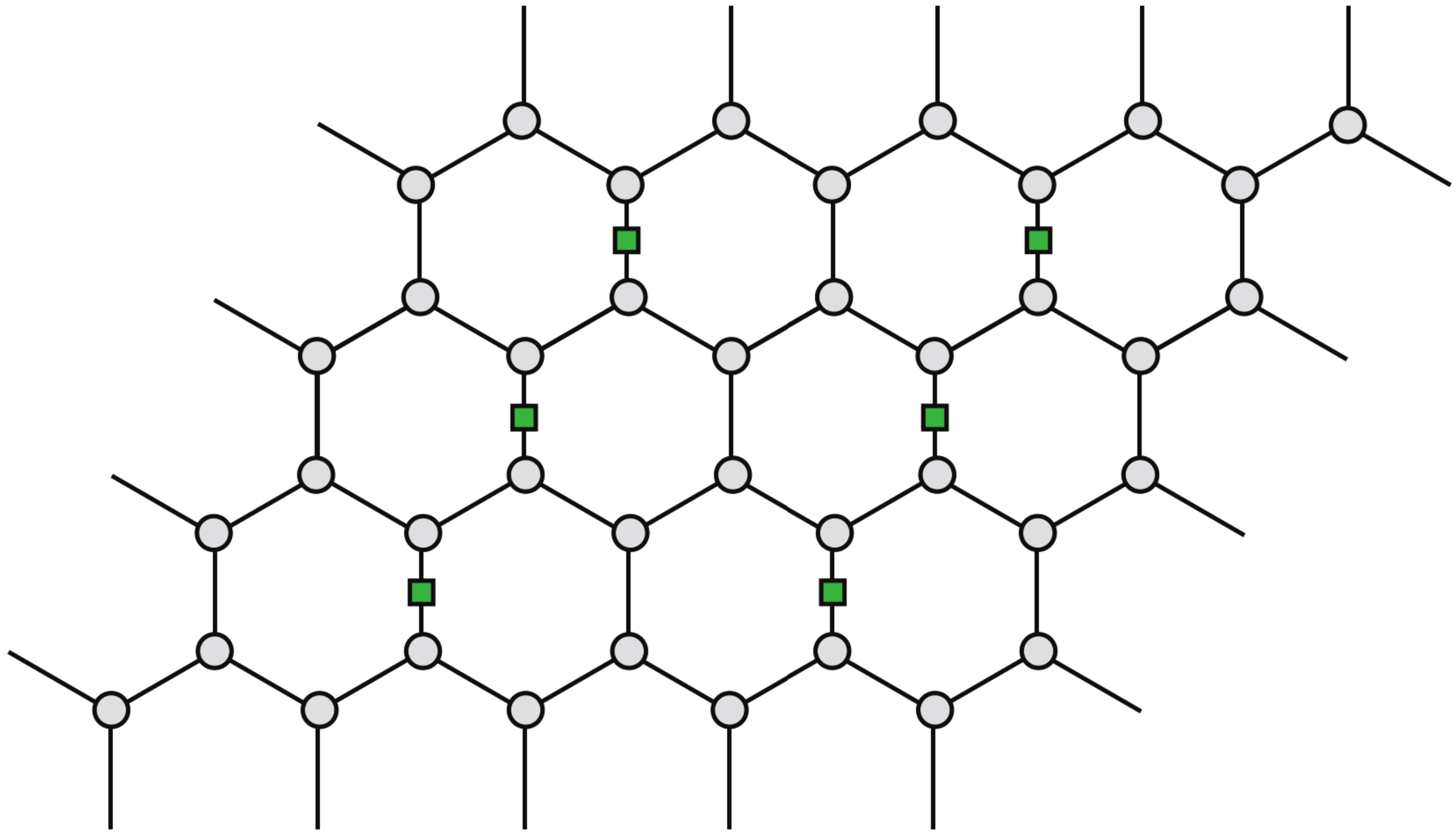} 
	\caption{ Schematic figure of the vortex-free\,(vortex-full) projector for the half-integer\,(integer) spin in the tensor network representation. Here, the gray circle with three legs stands for the local tensor $Q_{ijk}$, while the green square for $\sigma^z$. } 
	\label{fig:sector_change}
\end{figure}

On the other hand, the TN representation of the vortex-free projector for the half-integer spin is has a more complex structure. We define the following local tensor

\begin{align}
	Q_{000}	= \mathbb{I}_{2S+1},\quad
	Q_{011}	= -i U^x,\quad
	Q_{101}	= -i U^y,\quad
	Q_{110}	= -i U^z.
	\label{eq:Q_half_integer}
\end{align}
Here, we put the additional factor ``$-i$" except $Q_{000}$, which is essential to construct a projector in the case of half-integer spin. Note that the TN operator made of $Q_{ijk}$ without the additional factor is not a projector due to the relation $\{ U^\gamma, U^{\gamma'}\} = 0$. Furthermore, even with the factor, the resulting TN operator, say $\widetilde{Q}_{\rm LG}$, is not the vortex-free projector but the vortex-full projector, i.e., $W_p \widetilde{Q}_{\rm LG} = - \widetilde{Q}_{\rm LG}$ or $\widetilde{Q}_{\rm LG} = \prod_p (1-W_p)$. This can be easily verified using the following relation:

\begin{align}
	U^x Q_{ijk} = i\,v_{jj'} v^*_{kk'} Q_{ij'k'},\quad
	U^y Q_{ijk} = i\,v_{kk'} v^*_{ii'} Q_{i'jk'},\quad
	U^z Q_{ijk} = i\,v_{ii'} v^*_{jj'} Q_{i'j'k},
\end{align}
with

\begin{align}
	v = \begin{pmatrix}
		0 & i \\ 1 & 0
	\end{pmatrix}	
\end{align}
However, note that, in the current definition of the flux operator, the ground state of the $S=1/2$ Kitaev model is in the vortex-full sector: $W_p = U_1^x U_2^y U_3^z U_4^x U_5^y U_6^z = - \sigma_1^x \sigma_2^y \sigma_3^z \sigma_4^x \sigma_5^y \sigma_6^z$. Therefore, the $\widetilde{Q}_{\rm LG}$-operator is the desirable operator at least for the $S=\frac{1}{2}$ Kitaev model. By utilizing the $Z_2$ gauge redundancy, i.e.,

\begin{align}
	g_{ii'} g_{jj'} g_{kk'} Q_{i'j'k'} = Q_{i'j'k'}
\end{align}
with $g \in \{ \mathbb{I}_2, \sigma^z \}$, one can easily transform the vortex-full projector into the vortex-free projector by substituting the non-trivial element of the $Z_2$ group, i.e., $g=\sigma^z$, in the tensor network as depicted in Fig.\,\ref{fig:sector_change}. The green square stands for $g=\sigma^z$ that creates the vortices on two plaquettes sandwiching the bond. Note that the representation of the $Z_2$ invariant gauge group\,(IGG) applies to the case of the integer spin identically. Therefore, substituting the green squares into $Q_{\rm LG}$ of the integer spin as illustrated in Fig.\,\ref{fig:sector_change}, the resulting TN operator is the vortex-full projector\,\cite{HY19b} which is opposite to the case of the half-integer spin. In a similar way, utilizing the non-trivial element of $Z_2$ IGG, one can easily define not only those two, vortex-free and vortex-full, projectors but also a projector targeting arbitrary vortex sector with $D=2$ TN representation. 

\section{ The norm of the integer-spin loop gas state at the strong anisotropic point. }

\begin{figure}[h!]
	\includegraphics[width=0.98\textwidth]{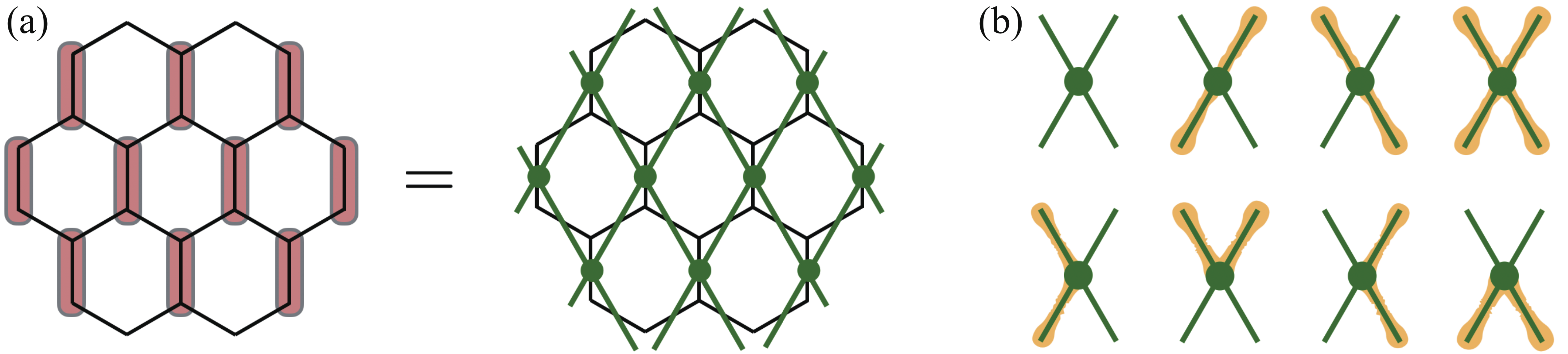} 
	\caption{ Schematic figure of (a) effective square lattice obtained by combining two sublattices\,(say $z$ or vertical bonds) and (b) eight different (partial) loop configurations on the vertex.} 
	\label{fig:square_lattice}
\end{figure}

In this section, we explicitly show that the loop gas state $|\psi^{\rm free}\rangle = Q_{\rm LG}^{\rm free} |\uparrow\rangle$ is identical to the product state $\bigotimes_{i=1}^{N_z} \frac{1}{\sqrt{2}} (|\tilde{\uparrow}_i\rangle + |\tilde{\downarrow}_i\rangle) $ , where $N_z$ is the number of $z$-bond, $|\tilde{\uparrow}_i\rangle = |\uparrow_i \uparrow_{i+\hat{z}}\rangle$ and $ |\tilde{\downarrow}_i\rangle = |\downarrow_i \downarrow_{i+\hat{z}}\rangle$. To this end, we first note that the number of the loop configurations on the honeycomb lattice is identical to the number of the configurations of the eight vertex model on an effective square lattice, $Z_{\rm 8ver}$. The square lattice is  obtained by combining two sublattices as depicted in Fig.\,\ref{fig:square_lattice}\,(a), and the eight different partial loop configurations are shown in Fig.\,\ref{fig:square_lattice}\,(b). Using $U^\alpha U^\beta \sim U^\gamma$ with $(\alpha,\beta,\gamma)$ being a permutation of $(x,y,z)$, one can show $(Q_{\rm LG}^{\rm free})^2 = Z_{\rm 8ver} \times Q_{\rm LG}^{\rm free}$. Therefore, the norm of the loop gas state is identical to the eight vertex model: $\langle \psi^{\rm free} | \psi^{\rm free} \rangle = Z_{\rm 8ver} \times \langle \uparrow | Q_{\rm LG}^{\rm free} |\uparrow\rangle = Z_{\rm 8ver}$. In the last equality, we use the fact that the overlap is zero if a loop configuration contains at least a single loop.

\section{Detailed analysis on the results of exact diagonalization and density matrix renormalization group }
In this section, we discuss the results of exact diagonalization\,(ED) and density matrix renormalization group\,(DMRG) methods.

\begin{figure}[h]
\includegraphics[width=0.8\textwidth]{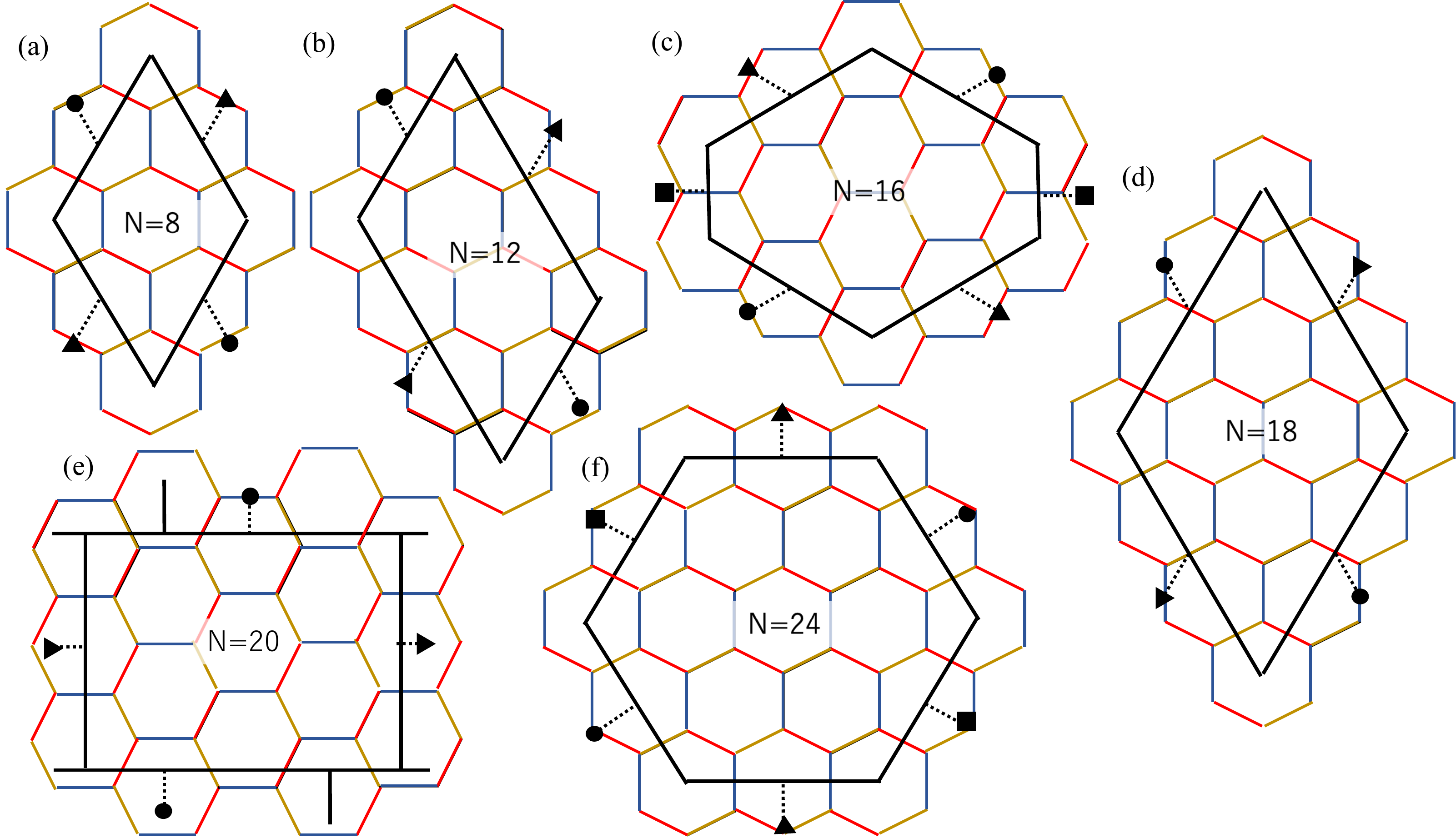}
\caption{(Color online) (a) $S=1$ Kitaev model on a honeycomb lattice up to $N=24$ cluster. Periodic boundary conditions are applied on the black lines with common symbols. Color of each bond corresponds to that in Fig.\,1\,(a) in the main text.} 
\label{fig:cluster}
\end{figure}
\begin{figure}[h]
\includegraphics[width=0.99\textwidth]{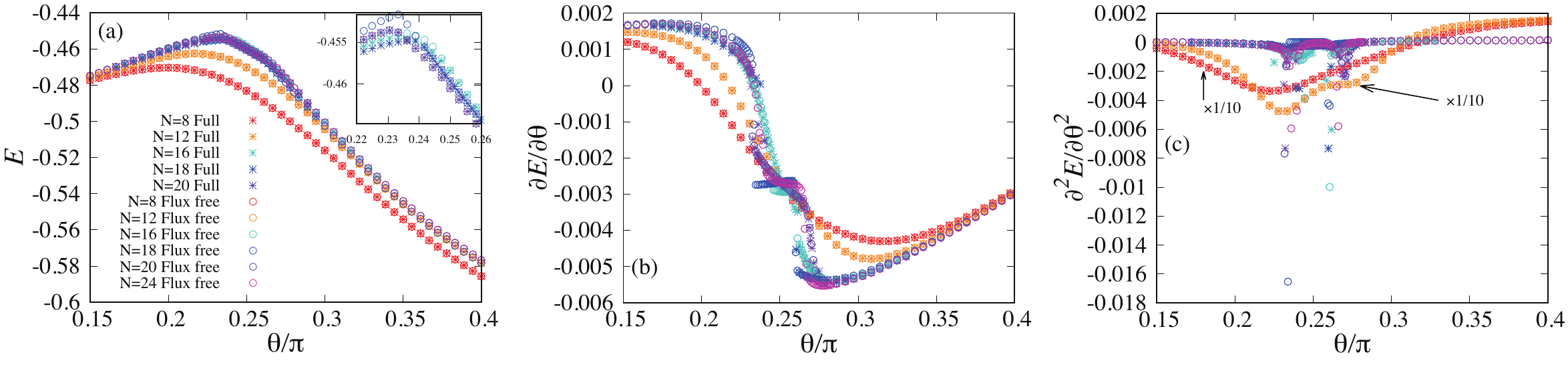}
\caption{(Color online) (a) The lowest energy energy per site, $E$, at $K_x=K_y$. Asterisk symbol denotes the ground state energy obtained by considering the full Hilbert space.
Open circle corresponds to the lowest energy in the flux free sector. The region around $\theta/\pi=0.24$ is magnified in inset. (b) $\partial E/\partial \theta$, and (c) $\partial^2 E/\partial \theta^2$ obtained with ED.}
\label{fig:ED1}
\end{figure}

We calculate the ground state energy per site, $E$, with ED. 
Since $W_p$ commutes the Hamiltonian (1), the Hilbert space of the Hamiltonian can be block diagonalized and classified into each space characterized by the set of the flux number on each hexagonal plaquette. 
Although this can reduce the computational cost, first we evaluate the ground state energy up to $N=20$ cluster without the block diagonalization by $W_p$.
Next, to see whether or not the ground state belongs to the flux free sector where $W_p=1$ is satisfied on all hexagons,  we investigate the lowest energy of the flux-free sector up to $N=24$ cluster. 
We summarize both results in Fig. \ref{fig:ED1}.

From Fig.\ref{fig:ED1} (a), we confirm that the ground state belongs to the flux-free sectors when the system size is large enough.
For the $N=16$ and $N=18$ clusters, the lowest energy of the flux-free sector coincides with the ground-state energy in $\theta/\pi \lessapprox 0.2$ and $0.24 \lessapprox \theta/\pi$, while a discrepancy exists in $0.2\lessapprox \theta/\pi \lessapprox 0.24$.  
This discrepancy exists in the dimer phase, not in the KSL phase.
We consider that this discrepancy is expected to be due to the system size effect.
Actually, for the $N=20$ cluster, the difference between the lowest energy of the flux free sector and the ground state energy becomes negligibly small in $0.2\lessapprox \theta/\pi \lessapprox 0.24$.
Thus, the ground state in the thermodynamic limit belongs to the flux free sector for $0 \le \theta/\pi \le 1/2$, which is consistent with the conjecture for the semi-classical model\cite{Baskaran08}.
Below, we focus on the results for the lowest energy of the flux free sector.

We find that the presence of the KSL phase at $\theta/\pi \approx 1/4$ becomes clear for the $N \ge 16$ clusters.
The ground state energy for the $N=16$ and $N=18$ clusters shows two cusps at $\theta_{c1}/\pi=0.235(5)$ and $\theta_{c2}/\pi =0.260(5)$, where the first derivative $\partial E/\partial \theta$ show a jump indicating the first-order transition.
In contrast, $E$ and $\partial E/\partial \theta$ for the $N=20$ and the $N=24$ cluster probably change continuously and $\partial E^2/\partial \theta^2$ shows local minima at $\theta_{c1}/\pi \approx 0.23$ and $\theta_{c2}/\pi \approx 0.27$.
Although it is difficult to conclude the order of the phase transition from the ED results for the small clusters, the KSL phase exists in the narrow region around $\theta/\pi=1/4$. 
From the ED results, we find that the spin-1 KSL phase at $K_x \approx K_y \approx K_z$ is quite narrow in comparison with that in the spin-1/2 model.
This is contrast to the spin-1/2 model, where the gapless KSL state\cite{Kitaev2006} at $K_x \approx K_y \approx K_z$ survives up to the chain limit, where two gapped KSL states also meet at $K_z=0$ and $K_x=K_y$.

\begin{figure*}[h]
\includegraphics[width=0.32\textwidth]{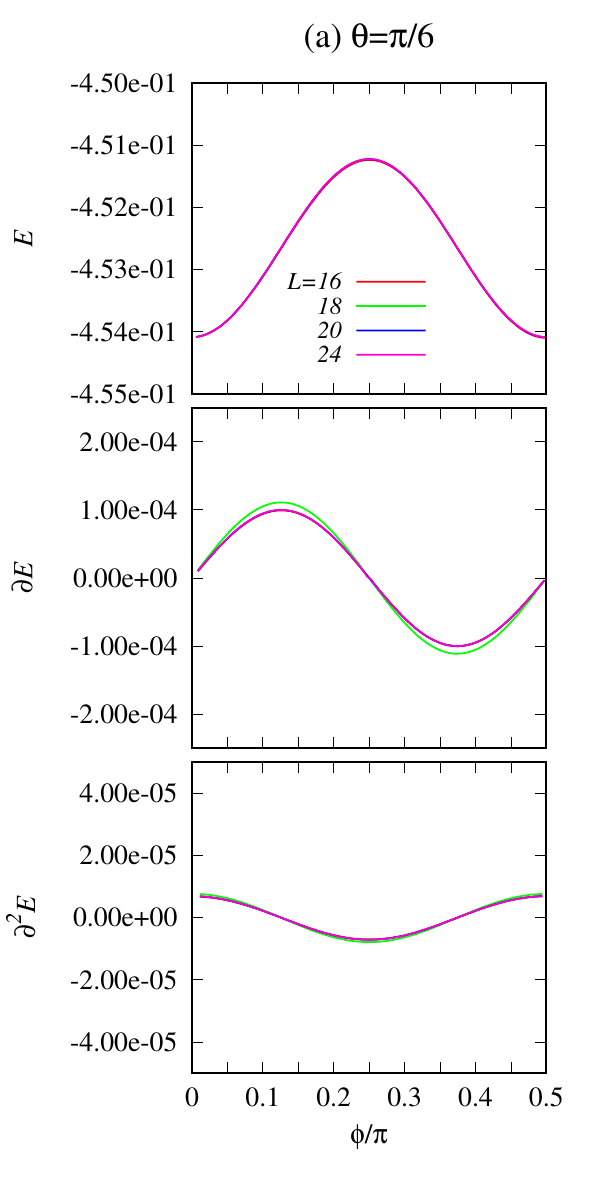}
\includegraphics[width=0.32\textwidth]{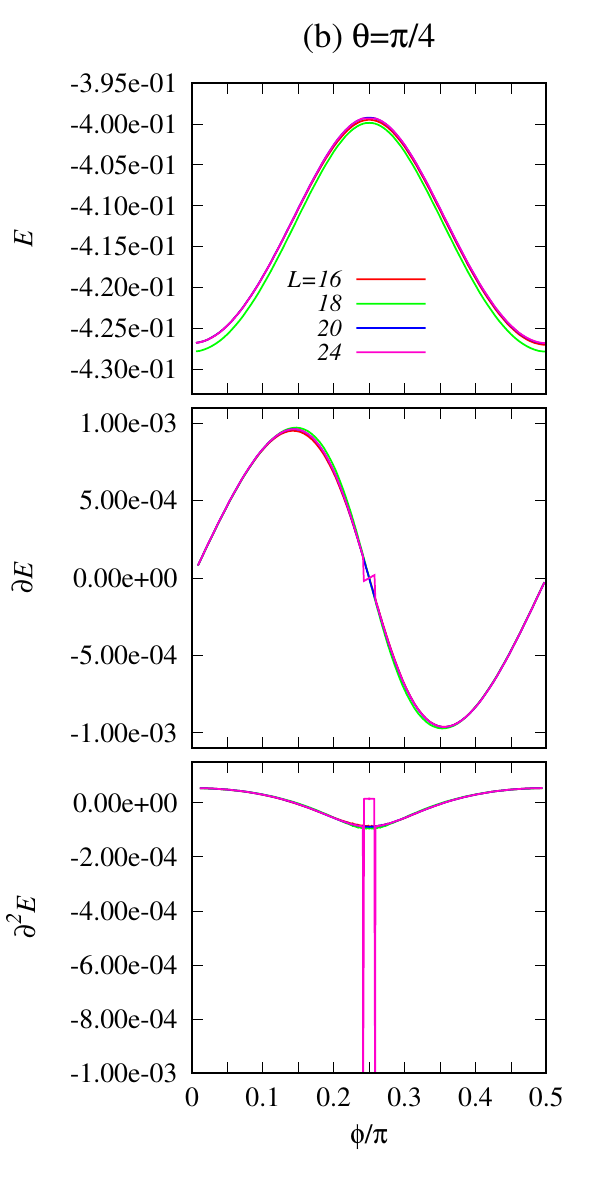}
\includegraphics[width=0.32\textwidth]{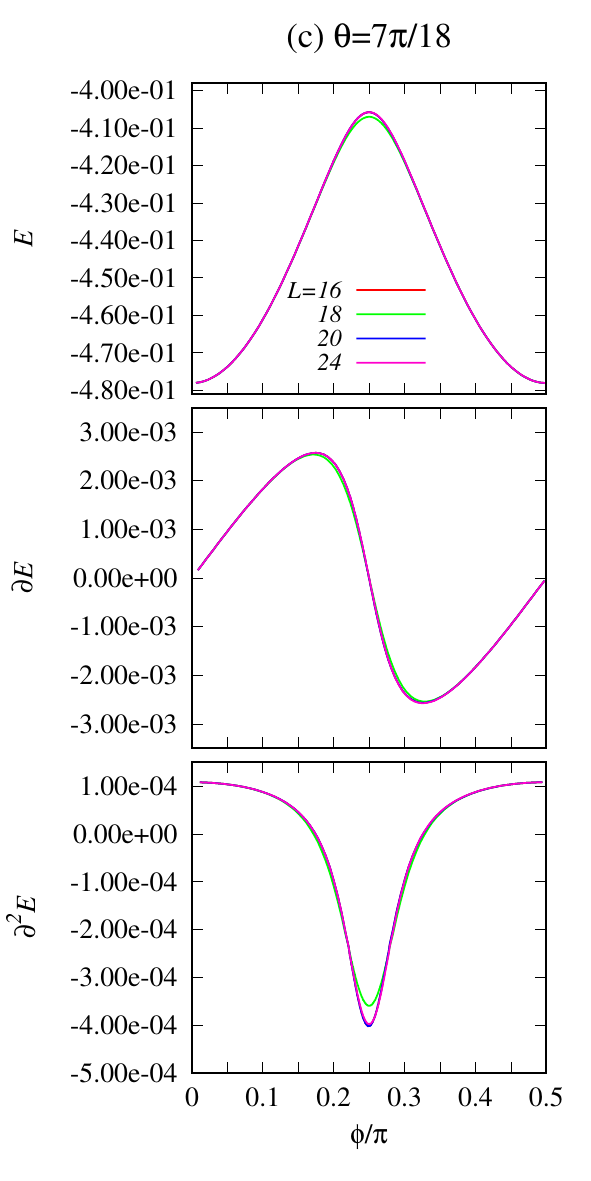}
\caption{(Color online) Typical behavior of $E$, $\partial E/\partial \phi$, and $\partial^2 E/\partial \phi^2$. $\theta/\pi=$(a)$1/6$, (b) $1/4$, and (c) $7/18$. }
\label{Ene2}
\end{figure*}

In the spin-1 model, the dimer state is stabilized in the limit, $K_z \gg K_x = K_y$.
This means that three dimer patterns can be competing by the amplitude of the three Kitaev interactions.
In the $S=1/2$ Kitaev model, those three dimer states, namely the gapped KSL states\cite{Kitaev2006}, are separated by the quantum phase transition.
To see whether such phase transition exists, we parameterize the anisotropy of the model as $K_x=\sin \theta \cos \phi$, $K_y=\sin \theta \sin \phi$, and $K_z=\cos \theta$, 
and calculate the $\phi$ dependence of the ground state energy at several $\theta$s.
Figure \ref{Ene2} shows the typical behavior of $E$ when $\phi$ changes.
For $\theta/\pi=1/4$, $\partial E/\partial \phi$ shows two jumps at $\phi \approx 0.24\pi$ and $\phi \approx 0.26\pi$ reflecting the KSL phase.
Except these two jumps, $E$ continuously changes without any divergence in the first derivative $\partial E/\partial \phi$ and the second derivative $\partial^2 E/\partial \phi^2$.
For $\theta/\pi=1/6$ and $7/6$, $E$ also changes continuously and show a maximum at $\phi/\pi=1/4$.
At $\phi/\pi=1/4$, $\partial^2 E/\partial \phi^2$ shows minimum, but the system-size dependence is small.
Thus, three dimer states appearing in the dimer limit are adiabatically connected each other.

\twocolumngrid

\begin{figure}[b!]
\includegraphics[width=0.48\textwidth]{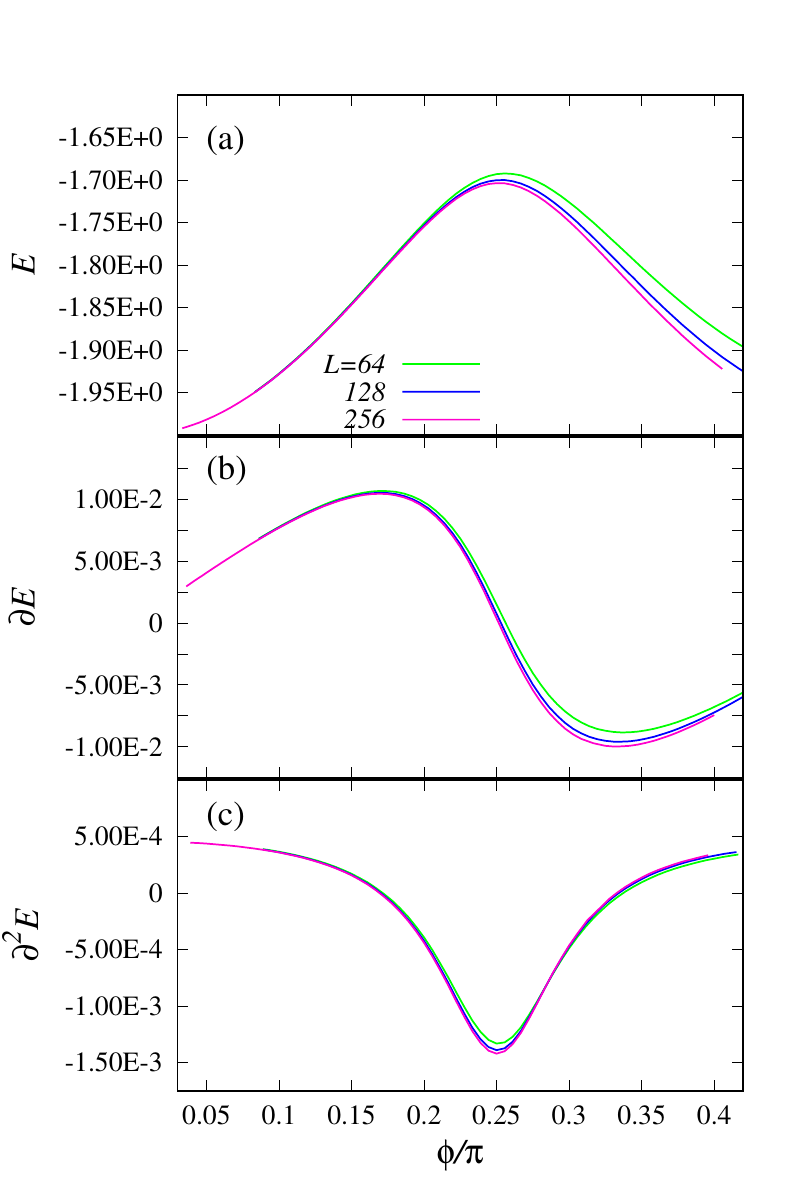}
\caption{(Color online) (a) Ground state energy $E$ at the chain limit, $K_z=0$. (b) The first derivative of $E$. (c) The second derivative of $E$.}
\label{DMRG_Ene}
\end{figure}

\begin{figure}[t!]
\includegraphics[width=0.48\textwidth]{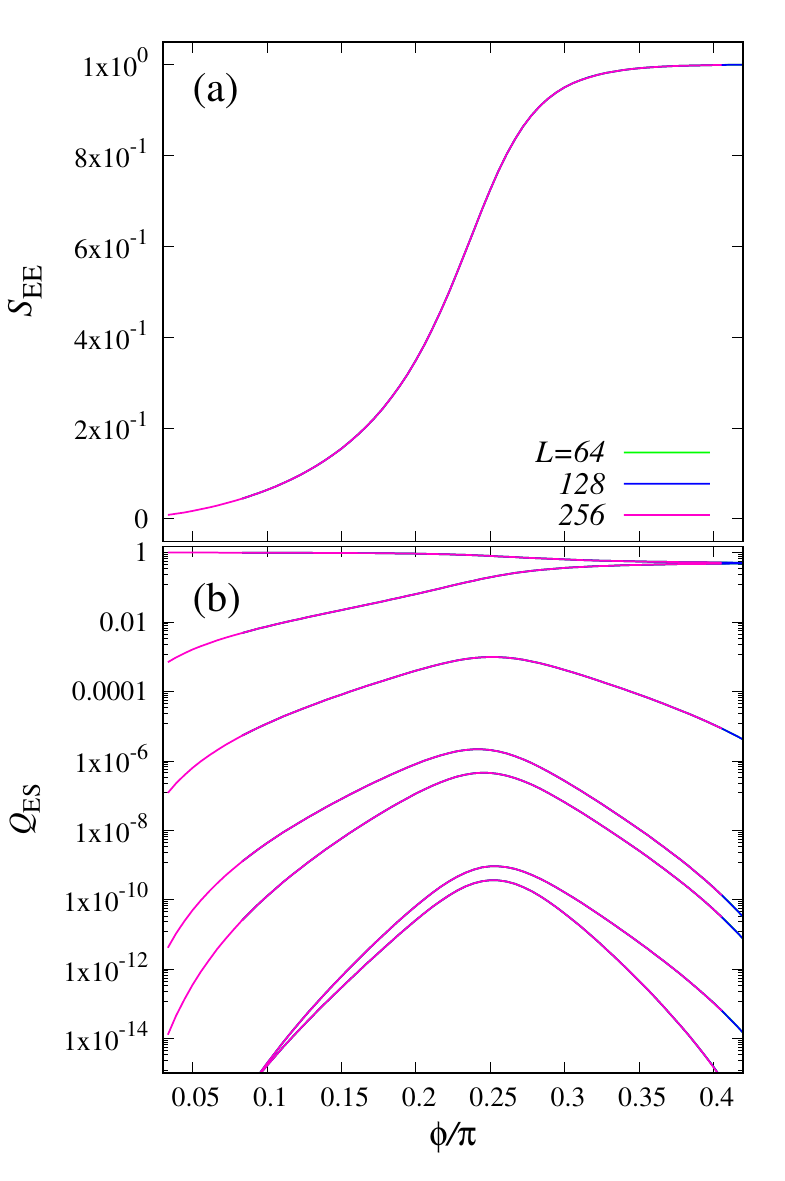}
\caption{(Color online) (a) The entanglement entropy, $S_{\rm EE}=\sum_i \lambda_i \log_2 \lambda_i$, where $\lambda_i$ is an eigenvalue of the reduced density matrix at $\theta=0$. (b) The entanglement spectrum, $Q_{\rm ES}$.}
\label{DMRG_EE}
\end{figure}

\onecolumngrid 

To clarify the above point, we calculate the ground state energy, entanglement entropy $S_{\rm EE}$, and the entanglement spectrum $Q_{\rm ES}$ at the chain limit with DMRG.
Note that $S_{\rm EE}=\lambda_i \sum_i \log_2 \lambda_i$, where $\lambda_i (i=0,1,2,\cdots)$ is the eigenvalue of the reduced density matrix $\rho$.
In DMRG calculation, we applied the open boundary condition to obtain highly accurate results.
The results are shown in Figs. \ref{DMRG_Ene} and \ref{DMRG_EE}.
The ground state energy $E$ in the chain limit changes continuously against $\phi$ accompanied by quite small system-size dependence of $\partial^2 E/\partial \phi^2$.
Indeed, the entanglement entropy $S_{\rm EE}$ changes continuously from zero at $\phi/\pi=0$ to unit at $\phi/\pi=1/2$, reflecting the fact that the dimer state on the $K_x$ bond gradually changes the dimer one on the $K_y$ bond.
Such continuous change of the state is also confirmed from the entanglement spectrum $Q_{\rm ES}$.
When the interaction for two spins located on the center of the system is absent, the system is perfectly divided into two parts.
At $\phi/\pi \approx 0$, the weakly interacting pairs are located on the center of the system. Therefore, the largest value of the eigenvalue of the density matrix, $\lambda_0$, is close to unit.
In contrast, when $\phi/\pi \approx 1/2$, two spins located on the center of the system are strongly interacting each other and construct the Ising ferromagnetic state with the doubly degeneracy.
This causes the doubly degeneracy of $\lambda_0$ and $\lambda_1$.
The obtained result indicate that in the $S=1$ Kitaev model, two isolated dimer states are adiabatically connected each other without the quantum phase transition.

\end{document}